\documentclass{llncs}
\usepackage[utf8]{inputenc}
\usepackage{makeidx}  % allows for indexgeneration
\usepackage{comment}
\usepackage{xcolor}
\usepackage{rotating}
\usepackage{graphicx}
\usepackage{subfigure}
\usepackage{amsmath}
\usepackage{acronym}
\usepackage{tabularx}
\usepackage{booktabs} %toprule midrule bottomrule
\usepackage{amssymb}
\usepackage{hyperref}

\newcommand{\Booking}{\textsf{Booking}}
\newcommand{\Amazon}{\textsf{Amazon}}
\renewcommand\tabcolsep{4pt}

\acrodef{NLP}{Natural Language Processing}
\acrodef{tf-idf}{term frequency–inverse document frequency}
\acrodef{tf}{term frequency}
\acrodef{idf}{inverse document frequency}
\acrodef{ts}{term specificity}

\begin{document}
\title{Mining Worse and Better Opinions} 
\subtitle{Unsupervised and Agnostic Aggregation\\ of Online Reviews\footnote{Extended version of ``Mining Worse and Better Opinions. Unsupervised and Agnostic Aggregation of Online Reviews'', to appear in Proc. of 17th International Web Engineering Conference (ICWE 2017).}}
%
%\titlerunning{Titlerunning}  % abbreviated title (for running head)
\author{Michela Fazzolari\inst{1} \and Marinella Petrocchi\inst{1} \and \\ Alessandro Tommasi\inst{2} \and Cesare Zavattari\inst{2}}
\authorrunning{M.Fazzolari et al.} % abbreviated author list (for running head)
%\tocauthor{Ivar Ekeland, Roger Temam, Jeffrey Dean, David Grove, Craig Chambers, Kim B. Bruce, and Elisa Bertino}
%
\institute{Institute of Informatics and Telematics (IIT-CNR), Pisa, Italy\\
\and
LUCENSE SCaRL, Lucca, Italy\\
\email{\{m.fazzolari,m.petrocchi\}@iit.cnr.it,\\ 
\{alessandro.tommasi,cesare.zavattari\}@lucense.it}
}
\maketitle              % typeset the title of the contribution
\begin{abstract}
%\notes{mari: altri suggerimenti per il titolo, lasciando comunque il sottotitolo:0) Speaking Highly or Speaking Ill? 1) Speaking More Highly. 2) Worse and Better Opinions. 3) Mining Worse and Better Opinions. 4) Mining High and Ill Opinions 5) alternative...?}
%\notes{mic: 5) For better or worse. (dovrebbe significare 'nel bene e nel male...')}
%\notes{mari: DOMANDA@lucensi: ma, teoricamente, potremmo anche comparare libri con hotel? tanto la terminologia e' calcolata sulla categoria..chissa' se funzionerebbe?}

%Online reviews represent an important resource for people to choose among multiple products and services. They also induce a powerful effect on customers' behaviour and, therefore, they undertake an influential role on the performance of business companies.

In this paper, we propose a novel approach for aggregating online reviews, according to the opinions they express. Our methodology is unsupervised - due to the fact that it does not rely on pre-labeled reviews - and it is agnostic - since it does not make any assumption about the domain or the language of the review content. 
%We do not adopt opinion mining techniques; rather, we propose a novel metric, measuring the adherence of the review content to the domain terminology extracted from the reviews set.
We measure the {\it adherence} of a review content to the domain terminology extracted from a review set. 
First, we demonstrate the informativeness of the adherence metric with respect to the score associated with a review. Then, we exploit the metric values to group reviews, according to the opinions they express. 
Our experimental campaign has been carried out on two large datasets 
%of hotel reviews and products reviews, 
collected from \Booking\ and \Amazon, respectively. 
%The results shown in this paper, other than being interesting on their own, also call for further investigations.

\end{abstract}
%%%%%%%%%%%%%%%%%%%%%%%%%%%%%%%%%%%%%%%%%%%%%%%%%%%%%%%%%
\section{Introduction}
%Online reviews are a powerful way to guide and influence the consumers' decision through online word-of-mouth. Relying on reviews, users may obtain an assessment of product quality, while companies acquire feedback to understand how their products and services are perceived. 
%Nowadays, consumers can easily and freely access information and exchange opinions about companies, services, products, by the means of online reviews. 

%With the popularity of online word-of-mouth activity, an increasing number of companies have started to offer online reviews services. For example, Amazon.com is well-known for its wide customer review system. From these reviews, consumers can obtain an assessment of product quality and be influenced in their purchase actions. At the same time, companies acquire immediate feedback to understand how their products and services are perceived, thus taking the opportunity to improve their products and services quality or to modify their marketing strategies.
Online reviews represent an important resource for people to choose among multiple products and services. They also induce a powerful effect on customers' behaviour and, therefore, they undertake an influential role on the performance of business companies.
Since the information available on reviews sites is often overwhelming, both consumers and companies benefit from effective techniques to automatically analysing the good disposition of the reviewers towards the target product. To this aim, opinion mining \cite{liu2012sentiment,pang2008opinion} deals with the computational treatment of polarity, sentiment, and subjectivity in texts.
%
%Opinion mining techniques \cite{liu2012sentiment} have been proved to enable the automatic analysis of large quantities of textual reviews, by extracting from them valuable information regarding the expressed opinions. 
However,  opinion mining is usually context-sensitive \cite{turney2002thumbs}, meaning that the accuracy of the sentiment classification can be influenced by the domain of the products to which it is applied \cite{ren2016context}. Furthermore, sentiment analysis may rely on annotated textual corpora, to appropriately train the sentiment classifier, see, e.g., \cite{esuli2006sentiwordnet}.
Also, most of the existing techniques are specialised for the English language: a cross-lingual adaptation is required in order to apply them to a different target language, \cite{lo2016amultilingual}.

\begin{comment}
Early approaches to this problem have focused on determining either the overall sentiment orientation (i.e., positive or negative) or the sentiment rating (e.g., one-to-five stars) of a review.
\end{comment}

%\notes{mic@all: CHECK from here}
In this paper, we propose an original approach to aggregate reviews with similar opinions. The approach is unsupervised, since it does not rely on labelled reviews and training phases. Moreover, it is agnostic, needing no previous knowledge on either the reviews domain or language. Grouping reviews is obtained by relying on a novel introduced metric, 
 called \textit{adherence}, which measures how much a review text inherits from a \textit{reference terminology}, automatically extracted from an unannotated reviews corpus. Leveraging an extensive experimental campaign over two large reviews datasets, in different languages, from \Booking\ and \Amazon\, we first demonstrate that the value of the adherence metric is informative, since it is correlated with the review score. Then, we exploit adherence to aggregate reviews according to the reviews positiveness. A further analysis on such groups highlights the most characteristic terms therein. This leads to the additional result of learning the best and worst features of a product.

 % of domain reviews. 
 %We used the score associated to each review only to prove the effectiveness of the proposed metric. Moreover, it is agnostic since we had no previous knowledge about the reviews domain nor about the language used in the reviews. We define a methodology that exploits this metric to cluster reviews, showing how these clusters correlate to the average review positiveness. Finally, we extract the most characteristic terms from each set of reviews, that give a flavour of better and worse features of a product.

%The remainder of the paper is as follows: 
In Section \ref{sec:adherence}, we define the adherence metric. Section \ref{sec:datasets} presents the datasets. Section \ref{ref:experiments} describes the experiments and their results. In Section \ref{sec:related}, we report on related work in the area.  Section \ref{sec:conclusions} concludes the paper.
\vspace{-0.3cm}
\section{Review Adherence to Typical Terminology}\label{sec:adherence}
%In this work, 
%we investigate the existence of a correlation between scores of reviews and the reviews' adherence to the domain terminology. 
We aim at proving that 
positive reviews - in contrast with negative ones - are generally more adherent to the emergent terminology of the whole review collection.  %whereas negative ones show more dissimilarity with respect to the emergent terminology in the whole review collection.
%As  a  side  benefit,  
This  will  provide  us  a  form  of  alternative  polarity
detection: indeed, we might estimate the relative polarity of a review by measuring
how adherent it is to the domain terminology. Because a meaningful comparison
against terminology requires a sizeable chunk of text, the proposed approach best applies to a
set of reviews. 
Here, we describe how the domain terminology is extracted and we define a measure of adherence of a piece of text against such terminology.
%This poses the grounds for the experiments described in Section \ref{ref:experiments}.

\vspace{-0.3cm}
\subsection{Extracting the Terminology}
\label{sec:terminology}
Every domain is characterized by key concepts, expressed by  a {\em domain terminology}: a set of terms that are either specific to the domain (e.g., part of its {\em jargon}, such as the term ``bluetooth" in the mobile domain) or that feature a specific meaning in the domain, uncommon outside of it (e.g., ``monotonous" in the math domain).
Identifying this terminology is important for two main reasons: i) avoiding that irrelevant terms (such as ``the", ``in", ``of" ...) have a weight in the computation of adherence; ii) knowing which key concepts are more relevant in a set of texts provides significant insight over their content.
\begin{comment}
\begin{enumerate}
    \item avoiding that irrelevant terms (i.e., functional words such as ``the", ``in", ``of" ...) have a weight in the computation of adherence;
    \item knowing which key concepts are more relevant in a set of texts (reviews, in our scenario) provides significant insight over their content.
\end{enumerate}
\end{comment}
The terminology is extracted in a domain and language agnostic way, with the 
%twofold 
benefit of not relying on domain and linguistic resources.
%and providing an easy way to apply the methodology.

{\em Contrastive approaches}~\cite{bonin2010acontrastive} to terminology extraction only rely on sets of raw texts in the desired language: i) a set belonging to the domain of interest and ii) a few others on generic topics (e.g., a collection of  books, newspaper articles, tweets -- easily obtainable, nowadays, from the Web). The contrastive approach work by comparing the characteristic frequency of the terms in the domain documents and in generic ones. The rationale is that generic, non-content words like ``the", as well as non specific words, will be almost equally frequent in all the available sets, whereas words with a relevance to the domain will feature there much more prominently than they do in generic texts.

There are many sophisticated ways to deal with multi-words, but any statistics-based approach needs to consider that, for $n$-grams\footnote{Constructions of $n$ words: ``president of the USA" is a 4-gram.} to be dealt with appropriately, the data needed scales up by orders of magnitude. 
%For our immediate purposes of evaluating adherence, 
For our purposes, we stick to the simpler form of single-term (or $1$-gram) terminology extraction.

%In Section \ref{sec:datasets}, we will provide details on which data we based the terminology extraction process on; we describe here the formal approach.

Let $\mathcal{D}$ be a set of documents belonging to the domain of interest $D$, and let $\mathcal{G}_1 \ldots \mathcal{G}_M$ be $M$ sets of other documents (the domain of each $\mathcal{G}_i$ is not necessarily known, but it is assumed not to be limited to $D$).
All terms occurring in documents of $\mathcal{D}$ ($T_\mathcal{D}$) as candidate members of $\mathbb{T}_\mathcal{D}$, the terminology extracted from $\mathcal{D}$.
For each term $t$, we define the {\em \ac{tf}} of a term $t$ in a generic set of documents $\mathcal{S}$ as:
\begin{equation}
    \text{tf}_{\mathcal{S}}(t) = \frac{|\{d \in \mathcal{S} | t \text{ occurs in } d\}|}{|\mathcal{S}|}
\end{equation}

\noindent (probability that, picking a document $d$ at random from $\mathcal{S}$, it contains $t$).
The \ac{tf} alone is not adequate to represent the meaningfulness of a term in a set of documents, since the most frequent words are non-content words\footnote{The ten most frequent words of the English language, as per Wikipedia (\url{https://en.wikipedia.org/wiki/Most_common_words_in_English}), are ``the", ``be", ``to", ``of", ``and", ``a", ``in", ``that", ``have", and ``I".}. Because of this, {\em \ac{idf}}~\cite{salton1988term} is often used to compare the frequency of a term in a document with respect to its frequency in the whole collection. In our setting, we can however simplify things, and just compare frequencies of a term inside and outside of the domain.
We do this by computing the {\em \ac{ts}} of a term $t$ over domain set $\mathcal{D}$ against all $\mathcal{G}_i$'s, which we define as:
\begin{equation}
    \text{ts}^\mathcal{D}_{\mathcal{G}}(t)=\frac{\text{tf}_{\mathcal{D}}(t)}{\displaystyle \min_{i=1..M}\text{tf}_{\mathcal{G}_i}(t)}    
\end{equation}

$\text{ts}^\mathcal{D}_{\mathcal{G}}(t)$ is effective at identifying very common words and words that are not specific to the domain (whose $\text{ts}$ will be close to 1), as well as words particularly frequent in the domain, with a $\text{ts}$ considerably higher than 1. Extremely rare words may cause issues: if $\mathcal{D}$ and $\mathcal{G}_i$'s are too small to effectively represent a term, such term will be discarded by default. We chose an empirical threshold $\theta_\text{freq}=0.005$, skipping all terms for which $\text{tf}_{\mathcal{D}}(t)<\theta_\text{freq}$. This value is
justified by the necessity to have enough documents per term, and 0.5\% is a reasonable figure given the size of our datasets.
We compute \ac{ts} for all $t\in T_{\mathcal{D}}$. We define:
\begin{equation}
    \mathbb{T}_\mathcal{D}=\{t|\text{ts}^\mathcal{D}_{\mathcal{G}}(t)\geq\theta_\text{cutoff}\}
\end{equation}

To set the value of $\theta_\text{cutoff}$, we might i) choose the number of words to keep (e.g., set the threshold so as to pick the highest relevant portion of $T_\mathcal{D}$) or ii) use an empirical value (higher than 1), indicating how much more frequent we ask a term to be, being a reliably part of the terminology. For our experiments, we have used this simpler alternative, empirically setting $\theta_\text{cutoff}=16$. Higher values include fewer terms in the terminology, improving precision vs.\ recall, whereas lower values include more terms, negatively affecting precision. This value was the one used in the experiments conducted in \cite{DBLP:journals/corr/VignaPTZT16}.
\vspace{-0.3cm}
\subsection{Adherence Definition}
%Upon extracting the terminology of a domain, 
The adherence ($\text{adh}$) of a document $d$ to a terminology $\mathbb{T}$ is defined as:
\begin{equation}
    \text{adh}_\mathbb{T}(d)=\frac{|\{t|t \text{ occurs in } d\}\cap\{t \in \mathbb{T}\}|}{|\{t|t \text{ occurs in } d\}|}
\end{equation}

It represents the fraction of terms in document $d$ that belongs to terminology $\mathbb{T}$. This value will typically be much smaller than 1, since a document is likely to contain plenty of non-content words, not part of the domain terminology. The specific value of adherence is however of little interest to us: we show how {\em more adherent} reviews tend to be more positive than those with lower values of adherence, only using the value for comparison, and not on an absolute scale.
\vspace{-0.2cm}

\section{Datasets}\label{sec:datasets}
%We present here the datasets used in our study.
%We used reviews as domain data for the terminology extraction as well as their associated score for the evaluation of the correlation between adherence and polarity, and general texts, used for the contrastive terminology extraction phase.

The first dataset consists of a collection of reviews from the \Booking\
%\footnote{\url{http://www.booking.com}}
website, during the period between June 2016 and August 2016. The second dataset includes reviews taken from the \Amazon\
%\footnote{\url{http://www.amazon.com}} 
website and it is a subset of the dataset available at \url{http://jmcauley.ucsd.edu/data/amazon}, previously used in \cite{mcauley2015inferring,mcauley2015image}. We also used a contrastive dataset to extract the domain terminology. 
%In the following we describe the characteristics of the three datasets.
\paragraph{\bf Booking Dataset.}%\label{sec:booking_dataset}
For the \Booking\ dataset, we had 1,135,493  reviews, related to 1,056 hotels in 6 cities. We only considered hotels with more than 1,000 reviews,  in any language. For each review, we focused on:
\begin{itemize}
%\item title: the title of the review;
\item score: a real value given by the reviewer to the hotel, in  the interval [2.5,10];
%\item author: the name of the reviewer;
%\item authorCountry: the reviewer location;
%\item date: the date of the review;
%\item tags: a list of tags associated with the review; 
\item posContent: a text describing the hotel pros;
\item negContent: a text describing the hotel cons;
\item hotelName: the name of the hotel which the review refers to.
\end{itemize}

As review text, we took the concatenation of posContent and negContent. 
%As previously stated, we considered 7 cities, namely London, Los Angeles, New York, Paris, Pisa, Rome and Sydney.
\paragraph{\bf Amazon Dataset.}
%We prepared the \Amazon\ dataset using the complete review data available at \url{http://jmcauley.ucsd.edu/data/amazon}.
Reviews in the \Amazon\ dataset are already divided according to the individual product categories. We chose two macro-categories, namely \textit{Cell Phones \& Accessories} and \textit{Health \& Personal Care} and we further selected reviews according to 6 product categories. For each review, we focused on:
%
%The original dataset includes reviews (with scores, text, degree of helpfulness of the review), product metadata (descriptions, category information, price, brand, and image features), and links (also viewed/also bought graphs). Among them, 
%
\begin{itemize}
    \item score: an integer assigned by the reviewer to the product (range [0,5]);
    \item reviewText: the textual content of the review;
    \item {\it asin}: the Amazon Standard Identification Number, that is a unique code of 10 letters and/or numbers that identifies a product. 
\end{itemize}

Table \ref{tab:datasets_stats} shows statistics extracted from the \Booking\ and the \Amazon\ dataset. %, described so far.

\begin{table}[htbp]
\centering
\caption{Outline of the datasets used in this study.}
\label{tab:datasets_stats}
\subtable[\Booking]{
\begin{tabular}{lrr}
\toprule
\textbf{City}  & \textbf{{\#}Hotels}   &\textbf{{\#}Rev.}   \\
\midrule
%Cape Town      &          &              \\               
%Dubai          &          &            \\       
London          &    358   &     521852       \\      
LosAngeles      &     57   &      51911         \\
NewYork         &    167   &     208917        \\
Paris           &    211   &     111103          \\
%Pisa            &     31   &      19713         \\            
%RioDeJaneiro   &          &             \\             
Rome            &    146   &      92321            \\        
Sydney          &     86   &     129676        \\        
\bottomrule
\end{tabular}
}
\hfill
\subtable[\Amazon]{
\begin{tabular}{lrr}
\toprule
\textbf{Product Category}  & \textbf{{\#}Prod.}   &\textbf{{\#}Rev.}   \\
\midrule
Bluetooth Headsets              &  937  & 124694  \\
Bluetooth Speakers              &   93  &  14941 \\  
Screen Protectors               & 2227  & 223007  \\        
Unlocked Cell Phones            & 1367  & 118889  \\     
%Appetite Control &  292  &  40246  \\      
Magnifiers                      &  210  &  12872 \\  
Oral Irrigators                 &   50  &  10768 \\       
\bottomrule
\end{tabular}
}
\end{table}

\paragraph{\bf Contrastive Terminology Dataset.}
In addition to the domain documents, originating from the above datasets, we used various datasets, collected for other purposes and projects, as generic examples of texts in the desired language, in order to extract the terminology as in Section~\ref{sec:terminology}. Table \ref{tab:contrastive} resumes the data used to construct the \textit{contrastive dataset}.
\begin{table}[b!]
\centering
\caption{Outline of the contrastive dataset.}
\begin{tabular}{lll}
\toprule
\textbf{English}                & \textbf{Italian}  & \textbf{French} \\
\midrule
220k online newspaper articles  & 1.28M forum posts & 198k tweets     \\
15.98M tweets                   & 7.37M tweets      &                 \\
\bottomrule
\end{tabular}
\label{tab:contrastive}
\end{table}

\vspace{-0.4cm}

%%%%%%%%%%%%%%%%%%%%%%%%%%%%%%%%%%%%%%%%%%%%%%%%%%%
\section{Experiments and Results}\label{ref:experiments}
%%%%%%%%%%%%%%%%%%%%%%%%%%%%%%%%%%%%%%%%%%%%%%%%%%%
%Before introducing experiments and results,  we give some useful notions. 
Each dataset $\mathcal{D}$ is organized in categories $\mathcal{C}_i$. Each category contains items that we represent by the set of their reviews $\mathcal{I}_j$. When performing experiments over $\mathcal{D}$, we extract the terminology of each category $\mathcal{C}_i$ ($\mathbb{T}_{\mathcal{C}_i}$). We then compute $\text{adh}_{\mathbb{T}_{\mathcal{C}_i}}(r)$ for each $r\in\mathcal{I}_j\in\mathcal{C}_i$ ($r$ is the single review).

%\notes{mic: cambiato equalization in balancing. Introdotta qui una spiegazione del perché il balancing riduce il bias e rimosso score class. Aggiunto sito web con additional material}

For the \Amazon\ dataset, $\mathcal{C}_i$ are the product categories, whereas $\mathcal{I}_j$'s are the products (represented by their sets of reviews). For the \Booking\ dataset, $\mathcal{C}_i$ are the hotel categories, whereas $\mathcal{I}_j$'s are the hotels (represented by their sets of reviews). 
We carried on experiments with and without {\it review balancing}. The latter has been considered to avoid bias: reviews with the highest scores are over-represented in the dataset, therefore the computation of the terminology can be biased towards positive terms. Thus, for each $\mathcal{C}_i$ and for each score, we randomly selected the same number of reviews.
%including some additional material associated to the paper
% setting $K_\text{bins}=3$. 
%%%%%%%%%%%%%%%%%%%%%%
\subsection{Adherence Informativeness}\label{sec:informativeness}
%%%%%%%%%%%%%%%%%%%%%%
A first analysis investigates if there exists a relation between the adherence metric - introduced in Section~\ref{sec:adherence} - and the score assigned to each review. 

\begin{figure}[htp]
\begin{minipage}[b]{0.48\textwidth}
\centering
\includegraphics[width=\textwidth]{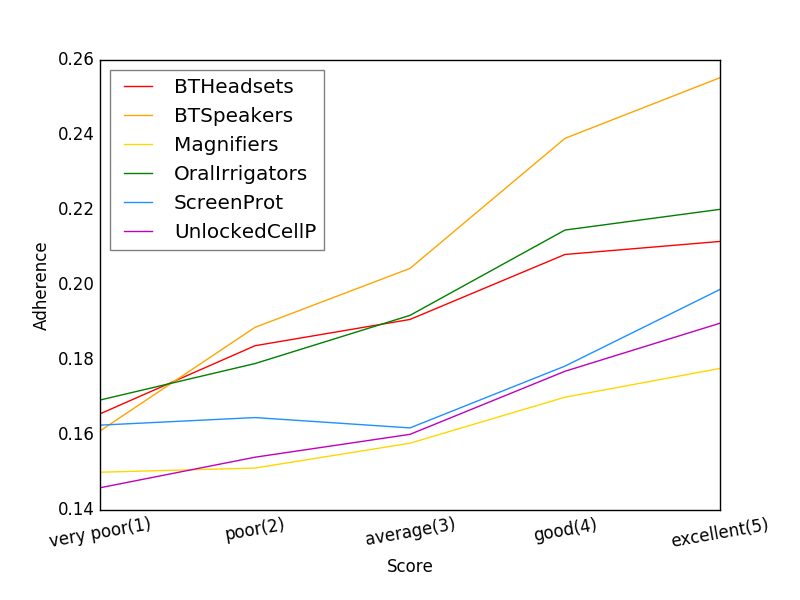}
\caption{Score vs adherence - \Amazon\ dataset - unbalanced.}
\label{fig:amazon_en_nobal_all}
\end{minipage}
\hspace{0.3cm}
\begin{minipage}[b]{0.48\textwidth}
\centering
\includegraphics[width=\textwidth]{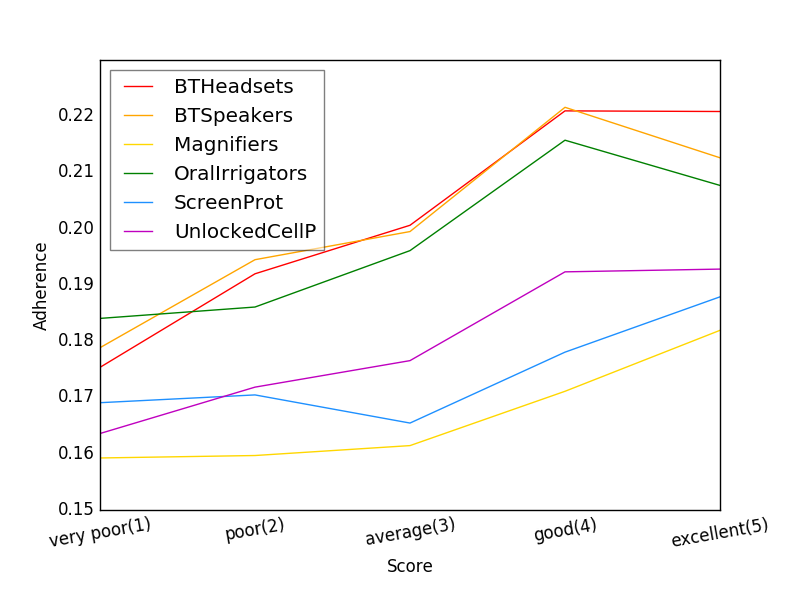}
\caption{Score vs adherence - \Amazon\ dataset - balanced.}
\label{fig:amazon_en_bal_all}
\end{minipage}
\end{figure}

{\bf Amazon Dataset.}
For each product category, we extract the reference terminology, by considering all the reviews belonging to that category against the contrastive dataset, for the appropriate language. Then, we compute the adherence value for each review. To show the results in a meaningful way, we grouped reviews in 5 bins, according to their score, and compute the average of the adherence values on each bin. The results are reported in Figure \ref{fig:amazon_en_nobal_all}.

\begin{comment}
%AMAZON NO BALANCED
\begin{figure}[htp]
\centering
\includegraphics[width=0.6\textwidth]{images/amazon_en_nobal_all}
\caption{Score vs adherence - \Amazon\ dataset - unbalanced.}
\label{fig:amazon_en_nobal_all}
\end{figure}
\end{comment}

The graph shows a line for each product category. Overall, it highlights that reviews with higher scores have higher adherence, implying a better correspondence with the reference terminology in comparison to reviews with lower scores. This result could be biased by the fact that reviews with higher scores are more represented in the dataset than the others. 

Therefore, we balanced the reviews in bins: we set $B$ as the number of reviews of the less populated bin and we randomly selected the same number of reviews from the other bins. Then, we recomputed the average adherence values and we obtained the results shown in Figure~\ref{fig:amazon_en_bal_all}, which confirm the trend of the previous graph. 

Even if the Bluetooth Speakers and Oral Irrigators categories feature a slight decreasing trend in the adherence value, when passing from reviews with score 4 to reviews with score 5, the general trend shows that the adherence metric is informative of the review score.

%AMAZON NOBALANCED AND BALANCED

{\bf Booking Dataset.}
In the second experiment, we group the hotel reviews accordingly to the city they refer to.  For each city, we extract the reference terminology and we compute the adherence value for each review.
To make the results comparable with the ones obtained for \Amazon\, we re-arrange the \Booking\ scoring system to generate a score evaluation over 5 bins. To this aim, we apply the score distribution suggested by \Booking\ itself, since \Booking\ scores are inflated to the top of the possible range of scores~\cite{mellinas2015}. Therefore, we consider the following bin distribution:
\begin{itemize}
    \item very poor: reviews with a score $\leq$ 3;
    \item poor: score $\in$ (3, 5];
    \item okay: score $\in$ (5, 7];
    \item good: score $\in$ (7, 9];
    \item excellent: score $>$ 9.
\end{itemize}

\begin{figure}[htbp]
\begin{minipage}[b]{0.48\textwidth}
\centering
\includegraphics[width=\textwidth]{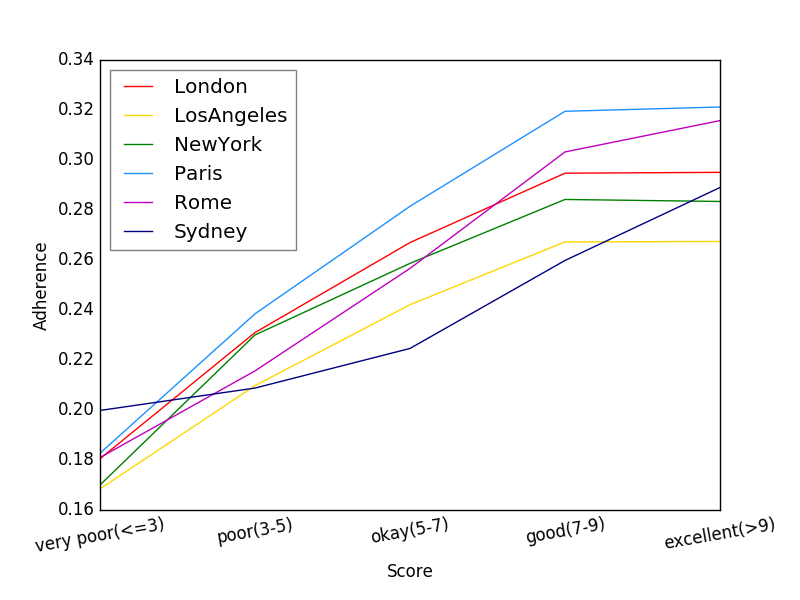}
\caption{Score vs adherence - \Booking\ - unbalanced.}
\label{fig:booking_en_nobal_all}
\end{minipage}
\hspace{0.3cm}
\begin{minipage}[b]{0.48\textwidth}
\centering
\includegraphics[width=\textwidth]{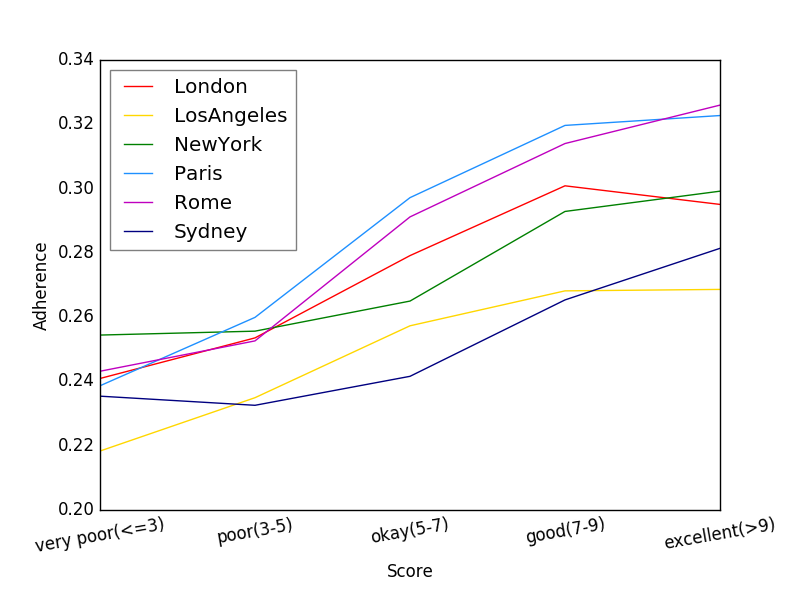}
\caption{Score vs adherence - \Booking\ - balanced.}
\label{fig:booking_en_bal_all}
\end{minipage}
\end{figure}

The results of the average adherence values on each bin are reported in Figure~\ref{fig:booking_en_nobal_all}. A line is drawn for each city, by connecting the points in correspondence to the adherence values. The graph suggests that the average adherence is higher for reviews with higher scores. Thus,  the higher the score of the hotel reviews, the more adherent the review to the reference terminology.

To avoid the bias caused by the over-representation of the positive reviews in the dataset, we compute the average adherence values by using a balanced number of reviews for each bin. The results, reported in Figure \ref{fig:booking_en_bal_all}, confirm the trend of the previous graph, even if the slope is smaller.

%BOOKING NOBALANCED AND BALANCED

\begin{comment}
\begin{figure}[htbp]
%AMAZON BALANCED - BOOKING BALANCED
\begin{minipage}[b]{0.48\textwidth}
\centering
\includegraphics[width=\textwidth]{images/amazon_en_bal_all}
\caption{Score vs adherence - \Amazon\ dataset - balanced.}
\label{fig:amazon_en_bal_all}

%\includegraphics[width=\textwidth]{images/booking_en_nobal_all}
%\caption{Score vs adherence - \Booking\ - unbalanced.}
%\label{fig:booking_en_nobal_all}
\end{minipage}
\hspace{0.3cm}
\begin{minipage}[b]{0.48\textwidth}
\centering
\includegraphics[width=\textwidth]{images/booking_en_bal_all}
\caption{Score vs adherence - \Booking\ dataset - balanced.}
\label{fig:booking_en_bal_all}
\end{minipage}
\end{figure}
\end{comment}

Working on balanced bins, the typical terminology of the reviews corpus has been recalculated, with respect to what computed for the original, unbalanced dataset. Thus, even considering the less populated bin, the average adherence value for the balanced dataset is not the same as the one in the unbalanced dataset for the same bin (this holds both for \Booking\ and \Amazon\ datasets).

The under-sampling applied to the majority classes to balance the reviews number leads to a deterioration of the results (both for \Amazon\ and \Booking\ - Figures \ref{fig:amazon_en_bal_all} and \ref{fig:booking_en_bal_all}.

\subsection{Good Opinions, Higher Adherence}\label{sec:goodhigher}
\label{subsec:posnegexp}

\begin{figure}[htbp]
\begin{minipage}[b]{0.48\textwidth}
\centering
\includegraphics[width=\textwidth]{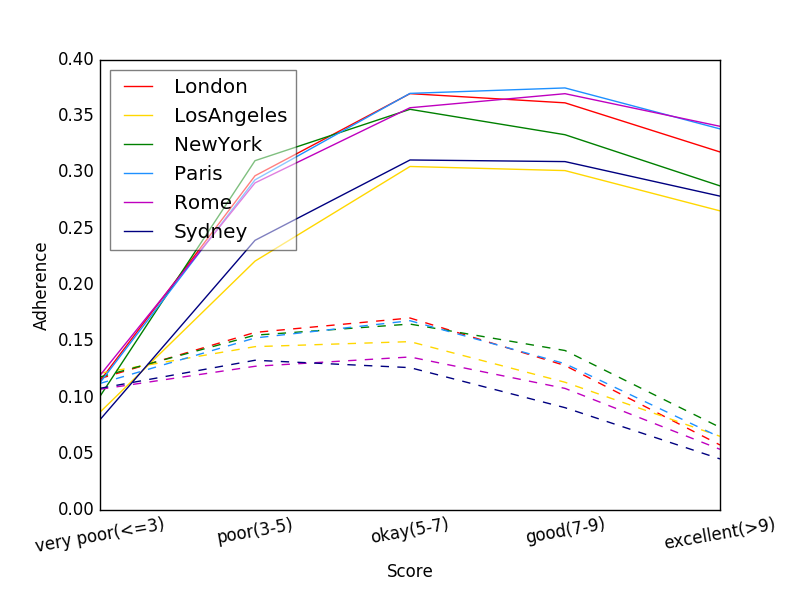}
\caption{Score vs adherence - \Booking\ unbalanced dataset - considering positive and negative contents separately.}
\label{fig:booking_en_nobal_posneg}
\end{minipage}
\hspace{0.3cm}
\begin{minipage}[b]{0.48\textwidth}
\centering
\includegraphics[width=\textwidth]{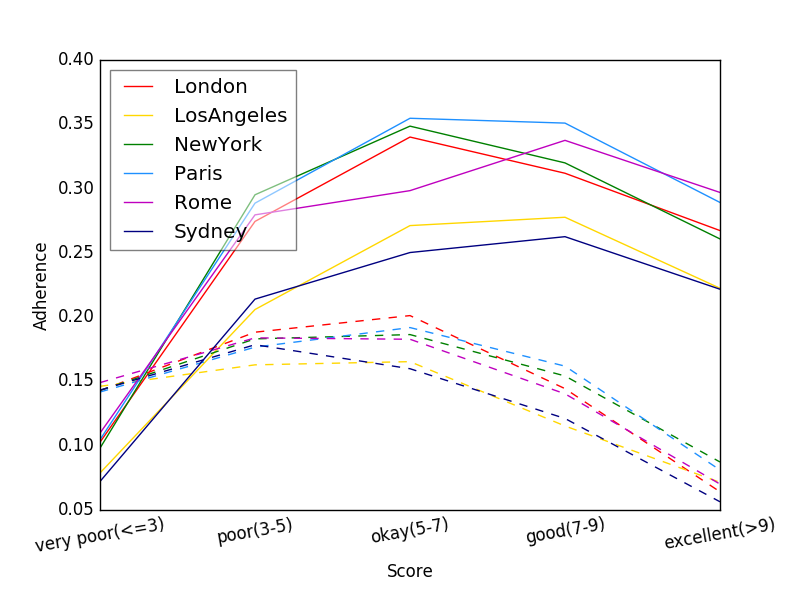}
\caption{Score vs adherence - \Booking\ balanced dataset - considering positive and negative contents separately.}
\label{fig:booking_en_bal_posneg}
\end{minipage}
\end{figure}

Interestingly, in the \Booking\ dataset, the text of each review is conveniently divided into positive and negative content. Thus, we perform an additional experiment, by only considering positive and negative chunks of reviews. For  each city, we group positive and negative contents of reviews and we compute the adherence value for each positive and negative chunk, with respect to the reference terminology. Finally, we average the adherence values according to the score bins. The results are reported in Figure \ref{fig:booking_en_nobal_posneg}, for the unbalanced dataset. In the graph, we report two lines for each city: the solid (dashed) lines are obtained by considering the positive (negative) contents of reviews. 
%while the dashed lines identify adherence values obtained by considering negative contents. 
The same colour for solid and dashed line corresponds to the same city. We also perform the same calculation by considering a balanced dataset (Figure \ref{fig:booking_en_bal_posneg}). 

Both the graphs highlight that there is a clear division between the solid and dashed lines. In particular, the average adherence obtained considering positive contents is, for most of the bins, above the average adherence computed considering negative contents. This separation is more evident when the review score increases (it does not hold for very poor scores). 
Overall, positive aspects of a hotel are described with a less varied language with respect to its negative aspects. Probably, this phenomenon occurs because unsatisfied reviewers  tend to explain what happened in details.

In addition to the average value, we also computed the standard deviation within each bin, that resulted to be quite high, as reported in Table \ref{tab:std_amazon_unbal}, Table \ref{tab:std_amazon_bal}, Table \ref{tab:std_booking_unbal} and Table \ref{tab:std_booking_bal}. This suggests that, even correlated with the score, the adherence is not a good measure when considering a single review, but its informativeness should be rather exploited by considering an ensemble of reviews, as detailed in Section \ref{sec:language-agnostic}.
\begin{table}[hbtp]
\centering
\caption{Average adherence and the average standard deviation for six Amazon categories - unbalanced dataset.}
\label{tab:std_amazon_unbal}
\footnotesize
\begin{tabular}{rrrrrrrrrrrrr}
\toprule
&
 \multicolumn{2}{c}{\textbf{Bluetooth}}&
 \multicolumn{2}{c}{\textbf{Bluetooth}}&
 \multicolumn{2}{c}{\textbf{Magnifiers}}&
 \multicolumn{2}{c}{\textbf{Oral}}&
 \multicolumn{2}{c}{\textbf{Screen}}&
 \multicolumn{2}{c}{\textbf{Unlocked}} \\
  &
 \multicolumn{2}{c}{\textbf{Headsets}}&
 \multicolumn{2}{c}{\textbf{Speakers}}&
 \multicolumn{2}{c}{}&
 \multicolumn{2}{c}{\textbf{Irrigators}}&
 \multicolumn{2}{c}{\textbf{Protectors}}&
 \multicolumn{2}{c}{\textbf{CellPhones}} \\
 \cmidrule(lr){2-3}
 \cmidrule(lr){4-5}
 \cmidrule(lr){6-7}
 \cmidrule(lr){8-9}
 \cmidrule(lr){10-11}
 \cmidrule(lr){12-13}

 \textbf{Bin}&
\textbf{Avg}&
\textbf{Std}&
\textbf{Avg}&
\textbf{Std}&
\textbf{Avg}&
\textbf{Std}&
\textbf{Avg}&
\textbf{Std}&
\textbf{Avg}&
\textbf{Std}&
\textbf{Avg}&
\textbf{Std}\\
 \cmidrule(lr){2-3}
 \cmidrule(lr){4-5}
 \cmidrule(lr){6-7}
 \cmidrule(lr){8-9}
 \cmidrule(lr){10-11}
 \cmidrule(lr){12-13}
1  	&0.17 	&0.07 	&0.16 	&0.08 	&0.15 	&0.08 	&0.17 	&0.07 	&0.16 	&0.07 	&0.15 	&0.06\\
2 	&0.18 	&0.07 	&0.19 	&0.07 	&0.15 	&0.07 	&0.18 	&0.07 	&0.16 	&0.07 	&0.15 	&0.06\\
3 	&0.19 	&0.07 	&0.20 	&0.08 	&0.16 	&0.08 	&0.19 	&0.07 	&0.16 	&0.07 	&0.16 	&0.07\\
4 	&0.21 	&0.08 	&0.24 	&0.09 	&0.17 	&0.08 	&0.21 	&0.08 	&0.18 	&0.08 	&0.18 	&0.08\\
5 	&0.21 	&0.09 	&0.26 	&0.11 	&0.18 	&0.09 	&0.22 	&0.08 	&0.20 	&0.09 	&0.19 	&0.10\\

\bottomrule
\end{tabular}
\end{table}
%%%%%%%%%%%%%%%%%%%%%%%%%%%%%%%%%%%%%%%%%%%%%%%%%%%%%%
\begin{table}[hbtp]
\centering
\caption{Average adherence and the average standard deviation for six Amazon categories - balanced dataset.}
\label{tab:std_amazon_bal}
\footnotesize
\begin{tabular}{rrrrrrrrrrrrr}
\toprule
&
 \multicolumn{2}{c}{\textbf{Bluetooth}}&
 \multicolumn{2}{c}{\textbf{Bluetooth}}&
 \multicolumn{2}{c}{\textbf{Magnifiers}}&
 \multicolumn{2}{c}{\textbf{Oral}}&
 \multicolumn{2}{c}{\textbf{Screen}}&
 \multicolumn{2}{c}{\textbf{Unlocked}} \\
  &
 \multicolumn{2}{c}{\textbf{Headsets}}&
 \multicolumn{2}{c}{\textbf{Speakers}}&
 \multicolumn{2}{c}{}&
 \multicolumn{2}{c}{\textbf{Irrigators}}&
 \multicolumn{2}{c}{\textbf{Protectors}}&
 \multicolumn{2}{c}{\textbf{CellPhones}} \\
 \cmidrule(lr){2-3}
 \cmidrule(lr){4-5}
 \cmidrule(lr){6-7}
 \cmidrule(lr){8-9}
 \cmidrule(lr){10-11}
 \cmidrule(lr){12-13}

 \textbf{Bin}&
\textbf{Avg}&
\textbf{Std}&
\textbf{Avg}&
\textbf{Std}&
\textbf{Avg}&
\textbf{Std}&
\textbf{Avg}&
\textbf{Std}&
\textbf{Avg}&
\textbf{Std}&
\textbf{Avg}&
\textbf{Std}\\
 \cmidrule(lr){2-3}
 \cmidrule(lr){4-5}
 \cmidrule(lr){6-7}
 \cmidrule(lr){8-9}
 \cmidrule(lr){10-11}
 \cmidrule(lr){12-13}
1	&	0.17	&	0.07	&	0.18	&	0.08	&	0.16	&	0.08	&	0.18	&	0.07	&	0.17	&	0.07	&	0.15	&	0.06	\\
2	&	0.19	&	0.07	&	0.2	&	0.07	&	0.16	&	0.08	&	0.19	&	0.06	&	0.17	&	0.07	&	0.16	&	0.06	\\
3	&	0.19	&	0.07	&	0.21	&	0.08	&	0.16	&	0.08	&	0.2	&	0.07	&	0.16	&	0.07	&	0.16	&	0.07	\\
4	&	0.21	&	0.08	&	0.23	&	0.08	&	0.17	&	0.08	&	0.22	&	0.08	&	0.18	&	0.08	&	0.18	&	0.08	\\
5	&	0.21	&	0.09	&	0.23	&	0.09	&	0.18	&	0.09	&	0.21	&	0.08	&	0.19	&	0.09	&	0.18	&	0.09	\\

\bottomrule
\end{tabular}
\end{table}
%%%%%%%%%%%%%%%%%%%%%%%%%%%%%%%%%%%%%%%%%%%%%%%%
\begin{table}[hbtp]
\centering
\caption{Average adherence and the average standard deviation for six Booking cities - unbalanced dataset.}
\label{tab:std_booking_unbal}
\footnotesize
\begin{tabular}{rrrrrrrrrrrrr}
\toprule
&
 \multicolumn{2}{c}{\textbf{London}}&
 \multicolumn{2}{c}{\textbf{LosAngeles}}&
 \multicolumn{2}{c}{\textbf{NewYork}}&
 \multicolumn{2}{c}{\textbf{Paris}}&
 \multicolumn{2}{c}{\textbf{Rome}}&
 \multicolumn{2}{c}{\textbf{Sydney}} \\

 \cmidrule(lr){2-3}
 \cmidrule(lr){4-5}
 \cmidrule(lr){6-7}
 \cmidrule(lr){8-9}
 \cmidrule(lr){10-11}
 \cmidrule(lr){12-13}

 \textbf{Bin}&
\textbf{Avg}&
\textbf{Std}&
\textbf{Avg}&
\textbf{Std}&
\textbf{Avg}&
\textbf{Std}&
\textbf{Avg}&
\textbf{Std}&
\textbf{Avg}&
\textbf{Std}&
\textbf{Avg}&
\textbf{Std}\\
 \cmidrule(lr){2-3}
 \cmidrule(lr){4-5}
 \cmidrule(lr){6-7}
 \cmidrule(lr){8-9}
 \cmidrule(lr){10-11}
 \cmidrule(lr){12-13}
1	&	0.18	&	0.15	&	0.17	&	0.13	&	0.17	&	0.15	&	0.18	&	0.14	&	0.18	&	0.14	&	0.20	&	0.16	\\
2	&	0.23	&	0.16	&	0.21	&	0.15	&	0.23	&	0.16	&	0.24	&	0.17	&	0.22	&	0.15	&	0.21	&	0.15	\\
3	&	0.27	&	0.17	&	0.24	&	0.16	&	0.26	&	0.17	&	0.28	&	0.18	&	0.26	&	0.17	&	0.22	&	0.15	\\
4	&	0.29	&	0.19	&	0.27	&	0.17	&	0.28	&	0.18	&	0.32	&	0.19	&	0.30	&	0.18	&	0.26	&	0.18	\\
5	&	0.30	&	0.19	&	0.27	&	0.18	&	0.28	&	0.18	&	0.32	&	0.19	&	0.32	&	0.18	&	0.29	&	0.19	\\

\bottomrule
\end{tabular}
\end{table}
%%%%%%%%%%%%%%%%%%%%%%%%%%%%%%%%%%%%%%%%%%%%%%
\begin{table}[hbtp]
\centering
\caption{Average adherence and the average standard deviation for six Booking cities - balanced dataset.}
\label{tab:std_booking_bal}
\footnotesize
\begin{tabular}{rrrrrrrrrrrrr}
\toprule
&
 \multicolumn{2}{c}{\textbf{London}}&
 \multicolumn{2}{c}{\textbf{LosAngeles}}&
 \multicolumn{2}{c}{\textbf{NewYork}}&
 \multicolumn{2}{c}{\textbf{Paris}}&
 \multicolumn{2}{c}{\textbf{Rome}}&
 \multicolumn{2}{c}{\textbf{Sydney}} \\

 \cmidrule(lr){2-3}
 \cmidrule(lr){4-5}
 \cmidrule(lr){6-7}
 \cmidrule(lr){8-9}
 \cmidrule(lr){10-11}
 \cmidrule(lr){12-13}

 \textbf{Bin}&
\textbf{Avg}&
\textbf{Std}&
\textbf{Avg}&
\textbf{Std}&
\textbf{Avg}&
\textbf{Std}&
\textbf{Avg}&
\textbf{Std}&
\textbf{Avg}&
\textbf{Std}&
\textbf{Avg}&
\textbf{Std}\\
 \cmidrule(lr){2-3}
 \cmidrule(lr){4-5}
 \cmidrule(lr){6-7}
 \cmidrule(lr){8-9}
 \cmidrule(lr){10-11}
 \cmidrule(lr){12-13}
1	&	0.24	&	0.18	&	0.22	&	0.15	&	0.25	&	0.22	&	0.24	&	0.18	&	0.24	&	0.17	&	0.27	&	0.23	\\
2	&	0.25	&	0.16	&	0.23	&	0.16	&	0.25	&	0.17	&	0.26	&	0.18	&	0.26	&	0.17	&	0.25	&	0.16	\\
3	&	0.27	&	0.17	&	0.25	&	0.16	&	0.27	&	0.17	&	0.29	&	0.18	&	0.29	&	0.16	&	0.23	&	0.14	\\
4	&	0.30	&	0.19	&	0.27	&	0.17	&	0.29	&	0.19	&	0.32	&	0.20	&	0.31	&	0.18	&	0.28	&	0.18	\\
5	&	0.30	&	0.20	&	0.28	&	0.19	&	0.29	&	0.19	&	0.31	&	0.18	&	0.32	&	0.19	&	0.30	&	0.21	\\

\bottomrule
\end{tabular}
\end{table}

\subsection{Extension to Different Languages}
\label{subsec:otherlanguages}

\begin{figure}[htp]
\centering
%\subfigure[Appetite control]
%{\includegraphics[width=0.49\textwidth]{images/avgdiff_appetite_control.png}}
%\hspace{2mm}
\subfigure[All content, Italian]
{\includegraphics[width=0.49\textwidth]{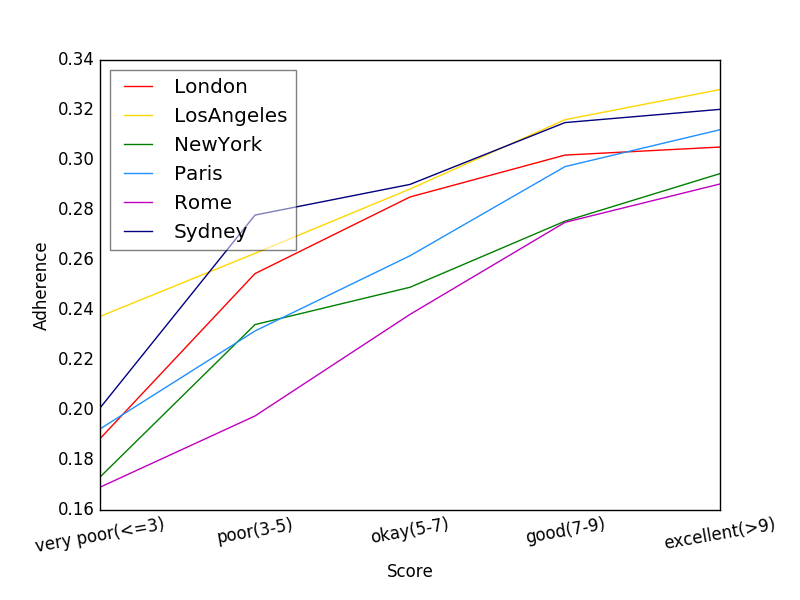}}
\subfigure[Positive/negative content, Italian]
{\includegraphics[width=0.49\textwidth]{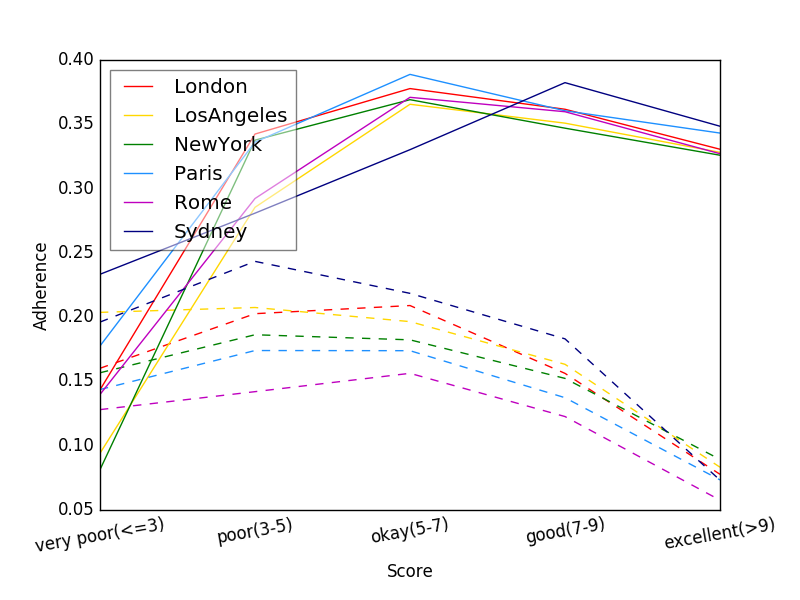}}
\subfigure[All content, French]
{\includegraphics[width=0.49\textwidth]{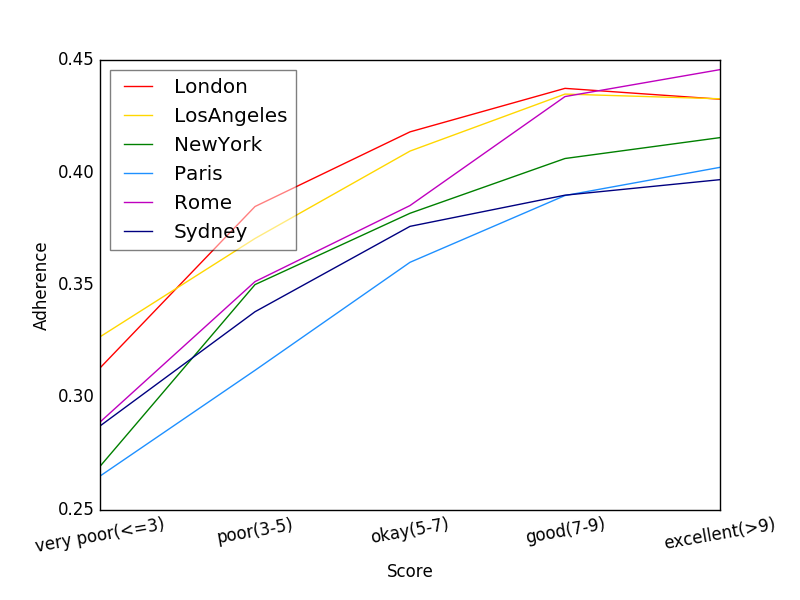}}
\subfigure[Positive/negative content, French]
{\includegraphics[width=0.49\textwidth]{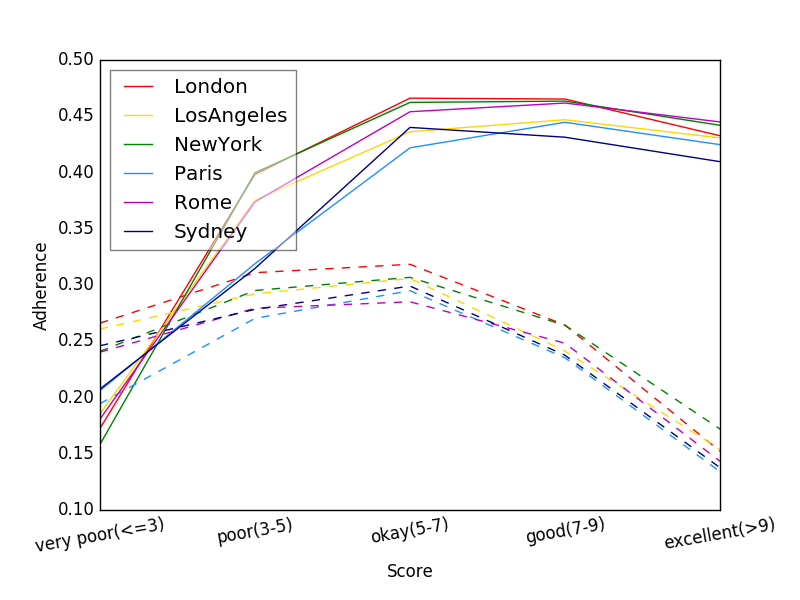}}
\caption{Score vs adherence - \Booking\  unbalanced  dataset - reviews in Italian and French.}
\label{fig:booking_nobal}
\end{figure}

The experiments described so far were realised by considering a subset of reviews in English, taken from the original \Booking\ dataset, which features other languages too. To further evaluate the informativness of the adherence metric, we selected two additional review subsets, in Italian and in French. For each subset, we drawn two graphs, the first considering all the reviews content, the second the separation between positive and negative contents. For these experiments, we considered imbalanced bins, due to the limited number of reviews available in each language. The results are reported in Figure \ref{fig:booking_nobal}. 
%\ref{fig:booking_it_nobal} for the Italian subset and in Figure \ref{fig:booking_fr_nobal} for the French subset, respectively.

\begin{comment}

\begin{figure}
\centering
\subfigure[All content]
{\includegraphics[width=0.49\textwidth]{images/booking_it_nobal_all}}
\subfigure[Positive/negative content]
{\includegraphics[width=0.49\textwidth]{images/booking_it_nobal_posneg}}
\caption{Score vs adherence - \Booking\  unbalanced  dataset - reviews in Italian.}
\label{fig:booking_it_nobal}
\end{figure}

\begin{figure}
\centering
\subfigure[All content]
{\includegraphics[width=0.49\textwidth]{images/booking_fr_nobal_all}}
\subfigure[Positive/negative content]
{\includegraphics[width=0.49\textwidth]{images/booking_fr_nobal_posneg}}
\caption{Score vs adherence - \Booking\ unbalanced  dataset - reviews in French.}
\label{fig:booking_fr_nobal}
\end{figure}

\end{comment}

Essentially, in both cases, it is confirmed that the higher the score, the higher the adherence when considering the overall text (Figure \ref{fig:booking_nobal}-a and Figure \ref{fig:booking_nobal}-c). Similarily, the graphs in Figure  \ref{fig:booking_nobal}-b and Figure \ref{fig:booking_nobal}-d show a clear division between positive and negative adherence values, when the score increases.

%%%%%%%%%%%%%%%%%%%%%%%%%%%%%%%%%%%%%%%%%%%%%%%%%%%%%%%%%%

\subsection{Language and Domain-Agnostic Reviews Aggregation}\label{sec:language-agnostic}
%\notes{Alternatives: language-independent or language-neutral or cross-language}

In this section, we present an application of the outcome found in previous ones. Given a set of texts, we propose to aggregate texts with positive polarity and texts with negative polarity, without a priori knowing the text language and domain, and without using any technique of \ac{NLP}, while exploiting only the adherence metric.
%defined in Section \ref{sec:adherence}.
%
We apply the following methodology: 
\begin{enumerate}
    \item For each review $r\in\mathcal{I}_j\in\mathcal{C}_i$ we compute the adherence $\text{adh}_{\mathbb{T}_{\mathcal{C}_i}}(r)$. 
    \item Reviews $r\in\mathcal{I}_j$ are sorted in ascending order {\it w.r.t.} their adherence value.
    \item Ordered reviews are split in bins with the same cardinality. We defined $K_\text{bins}$ bins, each holding $|\{r\in\mathcal{I}_j\}|/K_\text{bins}$ reviews in ascending order of adherence.
    \item For each bin $B_i$, we compute the average of the adherence value of the reviews it contains: $\text{Avg}_{\text{adh},i} = \frac{1}{R} \sum \text{adh}_{\mathbb{T}_{\mathcal{C}_i}}(r) $, as well as, for the purposes of validation, the average score provided by those reviews, $\text{Avg}_{\text{score},i} = \frac{1}{R} \sum \text{score}(r) $.
    \item Finally, we aim at proving that, when the average adherence value of each bin increases, the average score value also increases. Thus, we compute the percentage of $\mathcal{I}_j\in\mathcal{C}_i$ for which we observe:
    
    \begin{equation}\label{eq:first_last_equation}
    \text{Avg}_{\text{score},K_\text{bins}}  \geq \text{Avg}_{\text{score},1} 
    \end{equation}
    
    \begin{equation}\label{eq:monotonic_equation}
    \text{Avg}_{\text{score},i}  \geq \text{Avg}_{\text{score},i-1} 
    \end{equation}
    where $\text{Avg}_{\text{score},K_\text{bins}}$ is the average score for the last bin, $\text{Avg}_{\text{score},1}$ is the average score for the first bin, and $i = 1, \dots, K_\text{bins}$.
    
\end{enumerate}

Table~\ref{tab:amazon_first_last} reports the results for the \Amazon\ dataset. For each category $\mathcal{C}_i$, we apply the methodology three times, modifying the minimum number of reviews (\textit{minRev}) for each item $\mathcal{I}_j$, in order to discard items with few reviews. We set $K_\text{bins}$=3 and we report the number of items (\#$\mathcal{I}$) and the total number of reviews \textit{(\#Rev)} considered, plus the percentage of $\mathcal{I}_j\in\mathcal{C}_i$ for which \eqref{eq:first_last_equation} is true \textit{(\%)}. 
\begin{table}[hbtp]
\centering
\caption{\Amazon\ dataset - parameters: equation \eqref{eq:first_last_equation}, bins = 3.}
\label{tab:amazon_first_last}
\footnotesize
\begin{tabular}{lrrr|rrr|rrr}
\toprule
\textbf{Category $\mathcal{C}_i$} &\multicolumn{3}{c|}{\textbf{minRev=20}}& \multicolumn{3}{c|}{\textbf{minRev=50}} & \multicolumn{3}{c}{\textbf{minRev=100}} \\  
 
& {\#}$\mathcal{I}_j$ &\textit{{\#}Rev} & \textit{(\%)}   
& {\#}$\mathcal{I}_j$ &\textit{{\#}Rev} & \textit{(\%)}  
& {\#}$\mathcal{I}_j$ &\textit{{\#}Rev} & \textit{(\%)} \\
\midrule				
BluetoothHeadsets	& 817  & 108693 & 86 & 423 &  96393 & 93 & 223 &  82723 & 97 \\
BluetoothSpeakers	& 82   &  13155 & 96 &  54 &  12278 &100 &  27 &  10423 &100 \\
ScreenProtectors	& 1741 & 174320 & 83 & 781 & 144597 & 90 & 370 & 116337 & 96 \\	
UnlockedCellPhones	& 1116 &  97049 & 89 & 542 &  78836 & 94 & 257 &  58788 & 97 \\
%AppetiteControl     & 260  &  35862 & 85 & 130 &  31673 & 95 &  80 &  28090 & 97 \\	
Magnifiers          & 143  &   8763 & 72 &  46 &   5714 & 87 &  18 &   3694 &100 \\
OralIrrigators	    & 48   &  10301 & 85 &  32 &   9832 & 91 &  21 &   8987 & 90 \\
\bottomrule
\end{tabular}
\end{table}

This result shows that, considering 3 bins, the percentage of items for which the average score of the last bin is higher than the average score of the first bin is above 80\% for each category (except for \textit{Magnifiers} in case the minimum number of reviews is 20). Nevertheless, the percentage grows in almost all cases, when the minimum number of reviews increases. It exceeds 90\% for every category, when the minimum number of reviews is, at least, 100. 

Therefore, in the majority of cases, it is true that, when the average adherence of reviews belonging to the last bin is higher than the average adherence of reviews included in the first bin, the same relation exists between their correspondent average scores. 
This finding is mostly supported when we consider only items with many reviews (at least 100).

For the \Booking\ dataset, we straight consider only hotels with at least 100 reviews. We perform three experiments according to the languages of reviews (English, Italian, and French). For each experiment, $K_\text{bins}$=3 and we report the number of items (\#$\mathcal{I}$), the total number of reviews \textit{(\#Rev)} considered and the percentage of $\mathcal{I}_j\in\mathcal{C}_i$ for which \eqref{eq:first_last_equation} is true \textbf{(\%)}. The results are in Table \ref{tab:booking_first_last}. The percentage of items for which the average score of the last bin is higher than the average score of the first bin is above 90\% in all the cases. 
%\notes{mic@lucensi: riusciamo ad avere i risultati di sydney per tab 4 e tab 6? Altrimenti devo levarla dai grafici in sez 4.2 e 4.3}

\begin{table}[hbp]
\centering
\caption{\Booking\ dataset considering different languages - parameters: equation \eqref{eq:first_last_equation}, bins = 3. Not enough Italian reviews were available for Los Angeles and Sydney.}
\label{tab:booking_first_last}
\footnotesize
\begin{tabular}{lrrr|rrr|rrr}
\toprule
\textbf{Category $\mathcal{C}_i$} &\multicolumn{3}{c|}{\textbf{English}}& \multicolumn{3}{c|}{\textbf{Italian}} & \multicolumn{3}{c}{\textbf{French}} \\  
 
& {\#}$\mathcal{I}_j$ &\textit{{\#}Rev} & \textit{(\%)}   
& {\#}$\mathcal{I}_j$ &\textit{{\#}Rev} & \textit{(\%)}  
& {\#}$\mathcal{I}_j$ &\textit{{\#}Rev} & \textit{(\%)} \\
\midrule				
%Cape Town   &           &           &       &       &           &       &       &         &     \\
%Dubai	    &	267	    &	263351	&	99	&		&		    &		&		&		  &		\\
London	    &	356	    &	467863	&	97	&	76	&	11952	&	96	&	123	&	20507 &	94	\\
LosAngeles	&	56	    &	46700	&	93	&	-	&	-	    &	-	&	  7 &	  993 &	100	\\
NewYork	    &	163	    &	182438	&	95	&	27	&	6518	&	93	&	60	&	10753 &	90	\\
Paris	    &	211	    &	93164	&	96	&	6	&	806	    &	100	&	72	&	12623 &	90	\\
%Pisa	    &	211	    &	93164   &	95	&	28	&	4725	&	100	&	11	&	1553  &	100	\\
Rome	    &	144    	&	68543   &	97	&	64	&	11040	&	94	&	28	&	4197  &	93	\\
%RioDeJaneiro&           &           &       &       &           &       &       &         &     \\ 
Sydney      &    74     &   126744  &   100    & -      &    -       &  -     &  4  &   553   & 100    \\
\bottomrule
\end{tabular}
\end{table}

%According to previous findings, 

%We recall that we considered hotel with at least 100 reviews and three bins. 
%Therefore, this results confirm that when the average adherence computed on the last bin is higher than the average adherence computed on the first bin, the same occurs between the average scores of the correspondent bins.

Thus, given a set of reviews on, e.g., hotels, or restaurants, in any language,  we can identify a group of reviews that, on average, express better opinions than another group of reviews. Noticeably, this analysis works even if the associated score  is not available, i.e., it can be applied to general comments about items. 
%\medskip

We consider now if also relation \eqref{eq:monotonic_equation} is verified for each bin $i = 1, \dots, K_\text{bins}$, i.e., if the function between the ordered sets of average adherence values $\text{Avg}_{\text{adh},i}$ and average score values $\text{Avg}_{score,i}$ is a monotonic function.
By plotting the average score {\it vs} the average adherence, for some items, we found out a general upward trend. Nevertheless, there  were many spikes that prevent the function from being monotonic. Then, we tried to smooth down the curves by applying a moving average with $window=2$ and we then computed the percentage of $\mathcal{I}_j\in\mathcal{C}_i$ for which \eqref{eq:monotonic_equation} was verified. 
%As for the previous case, 
For the \Amazon\ dataset, we performed three experiments, modifying the minimum number of reviews required (\textit{minRev}) for each item, in order to discard items with few reviews. Results are in Table \ref{tab:amazon_monotonic}.

\begin{table}[htp]
\centering
\caption{\Amazon\ dataset - parameters: equation \eqref{eq:monotonic_equation}, bins = 3.}
\label{tab:amazon_monotonic}
\footnotesize
\begin{tabular}{lccc}
\toprule
\textbf{Category $\mathcal{C}_i$} & \textbf{minRev=20} (\%)& \textbf{minRev=50} (\%) & \textbf{minRev=100} (\%) \\  
\midrule				
BluetoothHeadsets	&	69	&	76	&	82	\\
BluetoothSpeakers	&	88	&	93	&	96	\\
UnlockedCellPhones	&	69	&	72	&	77	\\
%AppetiteControl	&	70	&	82	&	90	\\
Magnifiers	        &	58	&	72	&	72	\\
OralIrrigators	    &	77	&	81	&	76	\\
ScreenProtectors	&	58	&	66	&	73	\\
\bottomrule
\end{tabular}
\end{table}

Such results are worse with respect to the ones  in Table \ref{tab:amazon_first_last}. Nevertheless, in all cases (but \textit{Oral Irrigators}), the percentage values increase when $minRev$ increase (for \textit{Magnifiers}, it remains the same with $minRev = 50,100$). When $minRev = 100$, the percentage of $\mathcal{I}_j\in\mathcal{C}_i$ for which \eqref{eq:monotonic_equation} is true is above 72\%. % for all cases. 

For the \Booking\ dataset, due to the high number of available reviews, we also varied the number of bins from 3 to 5. We only considered reviews in English and computed the percentage of items for which the equation \eqref{eq:monotonic_equation} is true. Table \ref{tab:booking_monotonic} shows a clear degradation of performances when the number of bin  increases. 

\begin{table}[hbp]
\centering
\caption{\Booking\ dataset - parameters: equation \eqref{eq:monotonic_equation}.}
\label{tab:booking_monotonic}
\footnotesize
\begin{tabular}{lccc}
\toprule
\textbf{Category $\mathcal{C}_i$} & \textbf{bins=3 }(\%)& \textbf{bins=4} (\%) & \textbf{bins=5} (\%) \\  
\midrule				
%Dubai	&	92	&	74	&	57	\\
London	&	95	&	83	&	67	\\
LosAngeles	&	88	&	75	&	61	\\
New York	&	88	&	66	&	47	\\
Paris	    &	87	&	65	&	40	\\
%Pisa	&	97	&	69	&	54	\\
Rome	&	94	&	82	&	67	\\
Sydney  &   95   &  92     &    85   \\

\bottomrule
\end{tabular}
\end{table}

So far, the results indicate a relation between the increasing adherence values and the increasing score values. However, we cannot prove a strong correlation between adherence and score, either considering a single review or groups of reviews.
Therefore, we followed a different approach, by computing, for each item $\mathcal{I}_j \in \mathcal{C}_i$, the \textit{difference} between the average values of the first and last bin, both for the adherence and the score: 

$$
\Delta_{\text{adh}} (j) = \text{Avg}_{\text{adh},K_\text{bins}} - \text{Avg}_{\text{adh},1} 
$$
$$
\Delta_{\text{score}} (j) = \text{Avg}_{\text{score},K_\text{bins}}  - \text{Avg}_{\text{score},1} 
$$
%This values shown a significant correlation, therefore ??
If we average such differences for all the items $\mathcal{I}_j \in \mathcal{C}_i$, both for adherence and score, we obtain an average value for each category $\mathcal{C}_i$:

        \begin{equation}\label{eq:avg_diff_adh}
        \text{AvgD}_{\text{adh}} = \frac{1}{J} \sum_{j=1}^{J}  \Delta_{\text{adh}} (j) 
        \end{equation} 
        
        \begin{equation}\label{eq:avg_diff_score}
        \text{AvgD}_{\text{score}} = \frac{1}{J} \sum_{j=1}^{J} \Delta_{\text{score}} (j) 
        \end{equation}

where $J$ is the total number of items $j \in \mathcal{I}_j$. 

\begin{figure}[htp]
\centering
%\subfigure[Appetite control]
%{\includegraphics[width=0.49\textwidth]{images/avgdiff_appetite_control.png}}
%\hspace{2mm}
\subfigure[Bluetooth headsets]
{\includegraphics[width=0.49\textwidth]{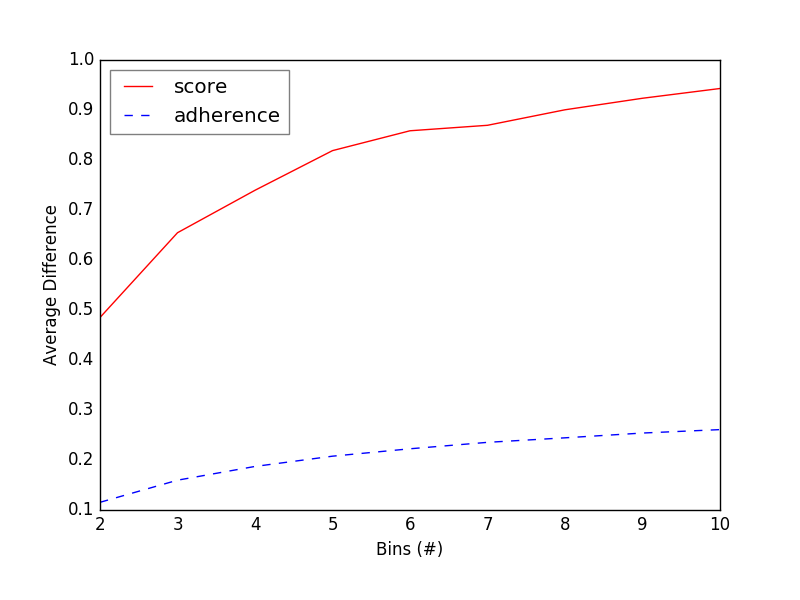}}
\subfigure[Bluetooth speakers]
{\includegraphics[width=0.49\textwidth]{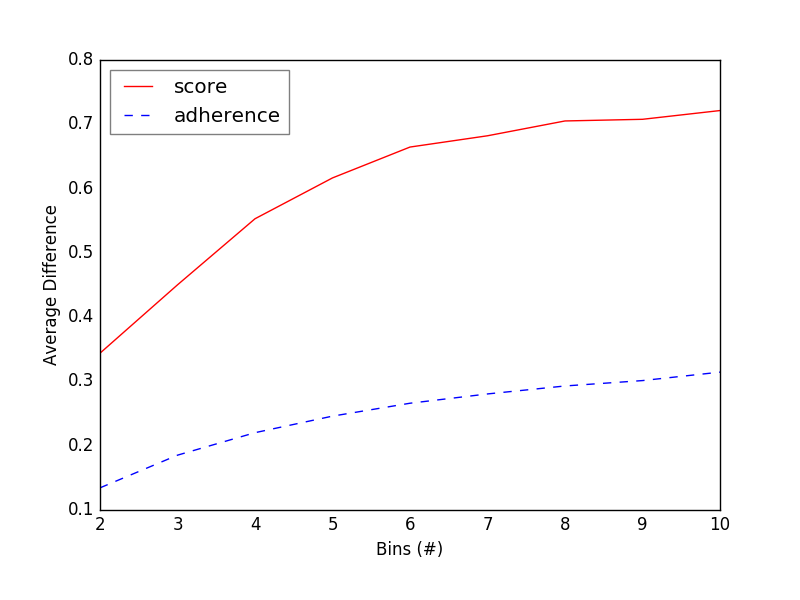}}
\subfigure[London (en)]
{\includegraphics[width=0.49\textwidth]{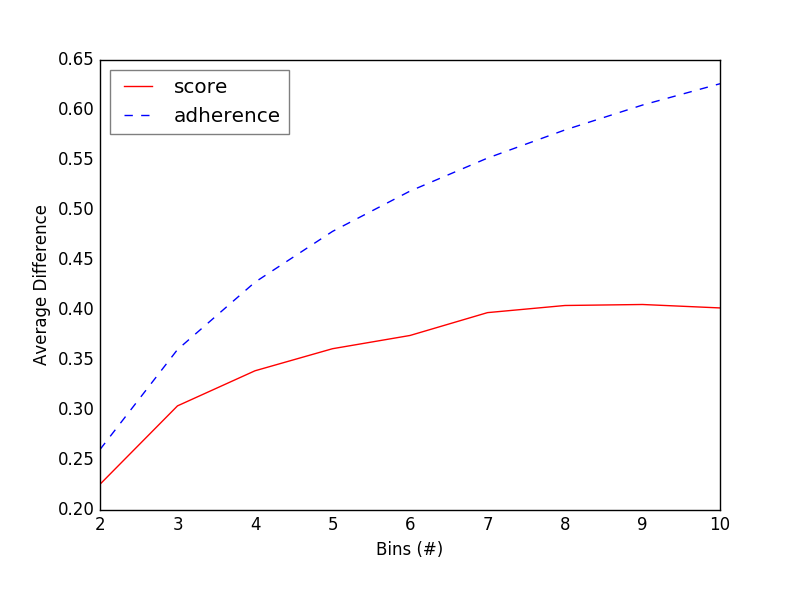}}
\subfigure[New York (en)]
{\includegraphics[width=0.49\textwidth]{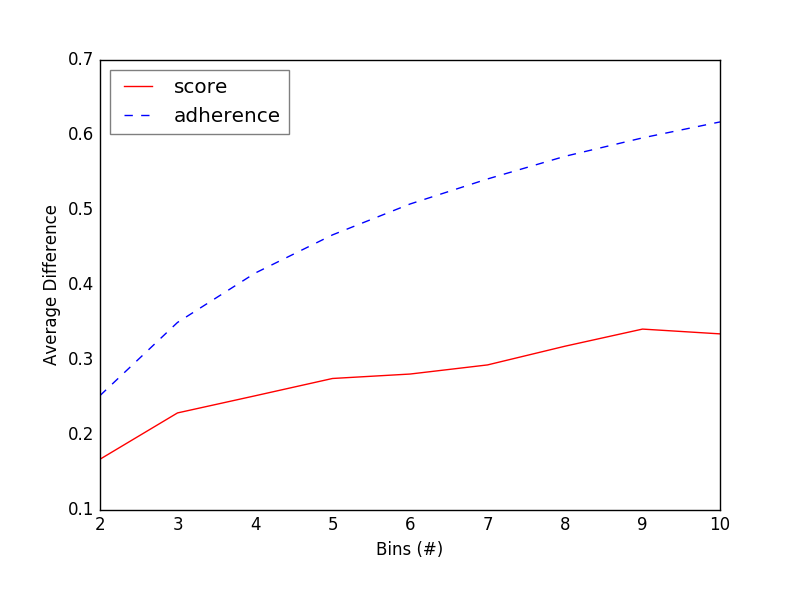}}
\caption{Average differences for \Amazon\ and \Booking\ example categories.}
\label{fig:differences}
\end{figure}

Figure~\ref{fig:differences} shows the values of \eqref{eq:avg_diff_adh} and \eqref{eq:avg_diff_score} for items belonging to some of the \Amazon\ and \Booking\ categories. In each graph, the x-axis reports the number of bins considered, wheres the y-axis represents the average differences values. We depicted the average differences for the adherence with a solid red line, while the average differences for the score with a dashed blue line. 
The graphs clearly show that, when the number of bin increases, the first and last bin include reviews which describe the product in a considerably different way, in term of positiveness. Thus, given a category, it is possible to discriminate among groups of reviews, related to that category, in such a way that each group expresses an opinion different from the others, ordered from the most negative to the most positive ones (or vice-versa).

%We computed the values in \eqref{eq:avg_diff_adh} and \eqref{eq:avg_diff_score} for items belonging to 2 categories of the \Amazon\ dataset and for items included in 2 categories of the \Booking\ dataset, varying the number of bins from 2 to 10. The results are reported in Figures \ref{fig:differences} for some of the \Amazon\ and \Booking\ categories, respectively.

\subsection{Representative Terms in First and Last Bins}
Here, we  extract the most recurrent terms in the positive and negative groups of  reviews. Given an item (either a hotel or a product), we consider the terms included in the positive set and in the negative set (last and first bins, with $K_\text{bins}=10$) that can be also found in the extracted terminology.

For each term, we compute the \ac{tf-idf} value (in this case, \ac{tf} is the term frequency inside the bin, that is the number of reviews that include such term), we sort the terms accordingly and we select the first 20 ones for the positive and negative set. We then remove the terms common to both sets, in order to identify the most discriminating ones. Table \ref{tab:amazon_terms} shows the terms extracted for two Amazon products.

\begin{table}
\caption{The most relevant positive and negative terms for two Amazon products.}
\label{tab:amazon_terms}
\centering
\begin{tabularx}{0.95\textwidth}{lXcXc}
\toprule
\textbf{Product} & \multicolumn{2}{c}{\textbf{Negative}} & \multicolumn{2}{c}{\textbf{Positive}} \\
\cmidrule(r){2-3}
\cmidrule(r){4-5}
        & \multicolumn{1}{l}{\textbf{Terms}}& \textbf{Score} & \multicolumn{1}{l}{\textbf{Terms}}& \textbf{Score}\\
\midrule
B005XA0DNQ  & refund, packaging, casing, disconnected, gift, battery, packaged, addition, hooked, plugging, shipping, 99, hook, speaker, purpose, sounds, kitchen & 2.9 &  compact, sound, great, retractable, portable, very, price, unbelievable, satisfied, product, easy, recommend, small, perfect, little, handy, size & 4.3 \\
\midrule
B0083RXA86  & stereo, impressive, mostly, charger, button, product, charging, switch, louder, useless, price, usb & 2.4 &  blue, charge, useful, very, coating, satisfied, quite, battery, enough, music, attachment, excellent, highly, quality, pleased & 4.6 \\
\bottomrule
\end{tabularx}
\end{table}

\begin{table}
\caption{The most relevant positive and negative terms for the Amazon categories.}
\label{tab:amazon_categ_terms}
\centering
\begin{tabularx}{0.95\textwidth}{lXX}
\toprule
\textbf{Product} & \textbf{Negative terms} & \textbf{Positive terms} \\
\midrule
BluetoothHeadsets  & charge, item, amazon, hear, purchased, device, problem, bought & fit, clear, recommend, highly, music, easy, excellent, comfortable, quality \\
\midrule
BluetoothSpeakers & reviews, speakers, gift, charge, volume, item, amazon, music, purchased, charging, bought & little, portable, excellent, bluetooth, easy, recommend, small, sounds, highly, quality, size\\
\midrule
ScreenProtectors & purchase, cover, bought, item, iphone, pack, instructions & perfect, samsung, clear, highly, fits, apply, perfectly \\
\midrule
UnlockedCellPhones & buy, phone, phones, seller, cell, everything, amazon, iphone, item, problem, bought & perfect, excellent, fast, card, easy, recommend, android, quality, sim\\
\midrule
AppetiteControl & try, tried, these, hungry, pills, reviews, products, oz, bottle, waste, eat, bought & loss, cambogia, lost, pounds, garcinia, recommend, diet, highly, definitely, lose, extract, exercise\\
\bottomrule
\end{tabularx}
\end{table}

After having extracted the terms for every item, we calculate their frequency within a whole category. Again, we remove the terms common to both sets (positive and negative), to highlight the most representative positive and negative terms for the category.
Table \ref{tab:amazon_categ_terms} shows the terms extracted for the five Amazon categories with more reviews.
Table \ref{tab:booking_categ_terms} shows the terms extracted for some Booking categories, for English, Italian and French. However, for Italian and French, few examples are shown, due to the low number of reviews for such datasets.

\begin{table}
\caption{The most relevant positive and negative terms for \Booking\ categories.}
\label{tab:booking_categ_terms}
\centering
\begin{tabularx}{0.99\textwidth}{lXX}
\toprule
\textbf{City} & \textbf{Negative terms} & \textbf{Positive terms} \\
\midrule
London \bf(en) & booked, bar, floor, wifi, stayed, stay, shower, receptionist, reception, booking, hotels & helpful, cleanliness, convenient, quiet, comfortable, beds, facilities, clean, excellent, friendly, size \\
\midrule
New York \bf(en) & booked, square, checked, floor, door, stayed, stay, reception, desk, reservation, booking, hotels & perfect, bathroom, helpful, cleanliness, comfortable, beds, clean, excellent, small, noisy, friendly, size \\
\midrule
Rome \bf(it) & albergo, prenotazione, servizi, stanze, hotel, struttura, soggiorno, reception, doccia, booking & disponibile, cortesia, gentile, confortevole, termini, ottima, gentilezza, abbondante, pulita, cordiale \\
\midrule
Paris \bf(fr) & emplacement, gare, accueil, anglais, chambres, hotel, clients, avons, sol, londres, wifi, lit, chambre, réservation, hôtel, réception, payé, bruit, booking & salle, géographique, déjeuner, proche, confortable, petite, petit, qualité, quartier, très, métro, calme, propreté, literie, situation, propre, agréable, proximité, bain \\
\bottomrule
\end{tabularx}
\end{table}

%We notice that the positive sets contain many terms (often adjectives) that can be expected in a positive context, such as excellent, clean, perfect, useful, satisfied, etc., confirming the intuition behind this work.

\section{Related Work}\label{sec:related}
\paragraph{Terminology extraction.} Automatic terminology extraction is one of the pillars of many terminology engineering processes, such as text mining, information retrieval, ontology learning, and semantic web technologies. The aim is to automatically identify relevant concepts (or terms) from a given domain-specific corpus. Within a collection of candidate domain-relevant terms, actual terms are separated from non-terms by using statistical and machine learning methods \cite{pazienza2005terminology}. Here, we rely on contrastive approaches, where the identification of relevant candidates is performed through inter-domain contrastive analysis \cite{penas2001corpus,chung2004identifying,basili2001contrastive}.

\paragraph{Opinion Mining.} Opinion mining techniques aim at automatically identifying polarities and sentiments in texts\cite{liu2012sentiment}, by, e.g., extracting subjective expressions representing personal opinions and speculations\cite{wilson2005opinion} or detecting so called contextual polarity of a word, i.e., the polarity acquired by the word contextually to the sentence in which it appears, see, e.g., \cite{wilson2005recognizing,wilson2009recognizing,muhammad2016contextual}. Often, opinion mining rely on lexicon-based approaches, involving the extraction of term polarities from sentiment lexicons and the aggregation of such scores to predict the overall sentiment of a piece of text, see, e.g., \cite{cambria2014senticnet,esuli2006sentiwordnet,cambria2015sentic,bravomarquez2016building}.

\paragraph{Clustering Opinions.} 
%There exist mainly supervised and unsupervised learning methods for polarity detection of text. 
Unsupervised learning does not require labelled data for the training process. Among them, 
clustering algorithms can be profitably used to find the natural clusters in the data, by calculating the distances or similarities from the centres of the clusters. 
Few efforts have been devoted to the study of polarity detection in online reviews with clustering techniques. In \cite{li2012application}, the authors present the clustering-based sentiment analysis approach, by applying three different techniques, namely TF-IDF weighting, voting mechanism and enhancement by hybrid with scoring method.
In \cite{ma2013acomparison,ma2015exploring}, the authors describe and experimental study of some common clustering techniques used for sentiment analysis of online reviews and investigate how any step of the clustering process (pre-processing, term weighting, clustering algorithm) can affect clustering results. 
%This study was extended in \cite{ma2015exploring} to document sentiment analysis.
%
The work in \cite{nagamma2015animproved} studies the relationship between online movie reviews and the box office incomes. The detection of sentiments is carried out by using \ac{tf} and \ac{idf} values as features and Fuzzy Clustering as algorithm. 

%In this paper, we cluster reviews with similar polarity. 
Here, we proposed an alternative approach to polarity detection, relying on automatic terminology extraction, in a domain and language agnostic fashion, and not relying on linguistic resources.

%Clustering algorithms have been also exploited in several sentiment analysis subtasks, such as topic clustering (\cite{andrzejewski2009latent}), product features clustering (\cite{zhai2011clustering}), sentiment clustering (\cite{zhu2014tripartite}), etc.

\vspace{-0.3cm}\section{Final Remarks}\label{sec:conclusions}
\vspace{-0.2cm}
%Clustering approaches can be useful for polarity detection due to the fact that they can provide accurate results without any human participation, linguist knowledge or training time. 
%
We presented a novel approach for aggregating  reviews, based on their polarity. The methodology did not require pre-labeled reviews and the knowledge of the reviews' domain and language. We introduced the adherence metric and we demonstrated its correlation with the review score. Lastly, we relied on adherence to successfully aggregate reviews, according to the opinions they express. 
\vspace{-0.4cm}

%\notes{SE avanza spazio (ahahah) potremmo inserire qualcosa su future works qui.}

\section*{\bf Acknowledgments.}\label{sec:Acknowledgments}
Research partly supported by MSCA-ITN-2015-ETN grant agreement \#675320 (\textit{European Network of Excellence in Cybersecurity}) and by Fondazione Cassa di Risparmio di Lucca, financing the project \textit{Reviewland}.

\bibliographystyle{splncs03}

%%%%%%%%%%%%%%%%%%%%%%%%%%%%%%%%%%%%%%%%%%%%%%%%%%%%%%%%%
\end{document}